\documentclass[preprint2]{emulateapj}
\usepackage{graphicx}
\usepackage{txfonts}
\usepackage{url}
\graphicspath{{./fig/}{./png/}}

\newcommand{\EQ}{\begin{equation}}
\newcommand{\EN}{\end{equation}}
\newcommand{\EQA}{\begin{eqnarray}}
\newcommand{\ENA}{\end{eqnarray}}
\newcommand{\eq}[1]{(\ref{#1})}
\newcommand{\EEq}[1]{Equation~(\ref{#1})}
\newcommand{\Eq}[1]{equation~(\ref{#1})}
\newcommand{\Eqs}[2]{equations~(\ref{#1}) and~(\ref{#2})}

\newcommand{\App}[1]{Appendix~\ref{#1}}
\newcommand{\Sec}[1]{\S\ref{#1}}

\newcommand{\Fig}[1]{Fig.~\ref{#1}}
\newcommand{\FFig}[1]{Figure~\ref{#1}}

\newcommand{\Figs}[2]{Figs~\ref{#1} and \ref{#2}}

\newcommand{\Figsss}[3]{Figs~\ref{#1}, \ref{#2} and \ref{#3}}

\newcommand{\bra}[1]{\langle #1\rangle}

\newcommand{\meanemf}{\overline{\mbox{\boldmath ${\cal E}$}}{}}{}
{}

{}
{}
{}
\newcommand{\meanEMF}{\overline{\mbox{\boldmath ${\cal E}$}}{}}{}
{}
{}
{}
{}
\newcommand{\meanAA}{\overline{\mbox{\boldmath $A$}}{}}{}
\newcommand{\meanBB}{\overline{\mbox{\boldmath $B$}}{}}{}
\newcommand{\meanJJ}{\overline{\mbox{\boldmath $J$}}{}}{}
\newcommand{\meanUU}{\overline{\mbox{\boldmath $U$}}{}}{}
{}
{}
\newcommand{\meanA}{\overline{A}}
\newcommand{\meanB}{\overline{B}}

\newcommand{\meanJ}{\overline{J}}

{}
{}
{}
{}

\newcommand{\eee}{\hat{\mbox{\boldmath $e$}} {}}

\newcommand{\xxx}{\hat{\mbox{\boldmath $x$}} {}}
\newcommand{\yyy}{\hat{\mbox{\boldmath $y$}} {}}

\newcommand{\gggg}{\mbox{\boldmath $g$} {}}

\newcommand{\kk}{\mbox{\boldmath $k$} {}}

\newcommand{\uu}{\mbox{\boldmath $u$} {}}
\newcommand{\UU}{\mbox{\boldmath $U$} {}}
\newcommand{\xx}{\mbox{\boldmath $x$} {}}
\newcommand{\aaa}{\mbox{\boldmath $a$} {}}
\newcommand{\bb}{\mbox{\boldmath $b$} {}}
\newcommand{\BB}{\mbox{\boldmath $B$} {}}

\newcommand{\jj}{\mbox{\boldmath $j$} {}}
\newcommand{\JJ}{\mbox{\boldmath $J$} {}}

\newcommand{\AAA}{\mbox{\boldmath $A$} {}}

\newcommand{\ff}{\mbox{\boldmath $f$} {}}
\newcommand{\hh}{\mbox{\boldmath $h$} {}}
\newcommand{\FF}{\mbox{\boldmath $F$} {}}

\newcommand{\nab}{\mbox{\boldmath $\nabla$} {}}
\newcommand{\OO}{\mbox{\boldmath $\Omega$} {}}

\newcommand{\kkappa}{\mbox{\boldmath $\kappa$} {}}

\newcommand{\SSSS}{\mbox{\boldmath ${\sf S}$} {}}

\newcommand{\emf}{\mbox{\boldmath ${\cal E}$} {}}

\newcommand{\ii}{{\rm i}}

\newcommand{\curl}{{\rm curl} \, {}}

\newcommand{\DDD}{{\cal D} {}}
\newcommand{\dd}{{\rm d} {}}
\newcommand{\const}{{\rm const}  {}}

\def\Co{\mbox{\rm Co}}

\def\Pm{\mbox{\rm Pr}_M}
\def\Rm{\mbox{\rm Re}_M}
\def\Rey{\mbox{\rm Re}}

\def\half{{\textstyle{1\over2}}}

\def\onethird{{\textstyle{1\over3}}}

\newcommand{\yaj}[3]{ #1, {AJ,} {#2}, #3}

\newcommand{\yapj}[3]{ #1, {ApJ,} {#2}, #3}

\newcommand{\yan}[3]{ #1, {Astron.\ Nachr.,} {#2}, #3}

\newcommand{\yana}[3]{ #1, {A\&A,} {#2}, #3}

\newcommand{\ygafd}[3]{ #1, {Geophys.\ Astrophys.\ Fluid Dyn.,} {#2}, #3}

\newcommand{\yjfm}[3]{ #1, {J.\ Fluid Mech.,} {#2}, #3}

\newcommand{\yprl}[3]{ #1, {Phys.\ Rev.\ Lett.,} {#2}, #3}

\newcommand{\ymn}[3]{ #1, {MNRAS,} {#2}, #3}
\newcommand{\ymonber}[3]{ #1, {Monatsber.\ Dtsch.\ Akad.\ Wiss.,} {#2}, #3}
\newcommand{\ynat}[3]{ #1, {Nature,} {#2}, #3}

\newcommand{\ypre}[3]{ #1, {Phys.\ Rev.\ E,} {#2}, #3}

\newcommand{\yjour}[4]{ #1, {#2}, {#3}, #4}

\newcommand{\ybook}[3]{ #1, {#2} (#3)}

\newcommand{\pana}[1]{ #1, {A\&A} (to be published)}

\newcommand{\pmn}[1]{ #1, {MNRAS} (in press)}

\submitted{}
\begin{document}

\title{Magnetic diffusivity tensor and dynamo effects in rotating
and shearing turbulence}
\author{A. Brandenburg\altaffilmark{1}, K.-H. R\"adler\altaffilmark{2},
M. Rheinhardt\altaffilmark{2}, and P. J. K\"apyl\"a\altaffilmark{1,3}}

\altaffiltext{1}{
NORDITA, Roslagstullsbacken 23, SE-10691 Stockholm, Sweden
}\altaffiltext{2}{
Astrophysical Institute Potsdam, An der Sternwarte 16, D-14482 Potsdam, Germany
}\altaffiltext{3}{
Observatory, University of Helsinki, PO Box 14, FI-00014 University of Helsinki, Finland
\\ $ $Revision: 1.171 $ $ (\today)
}

\begin{abstract}
The turbulent magnetic diffusivity tensor is determined in the presence
of rotation or shear. The question is addressed whether dynamo action
from the shear--current effect can explain large-scale magnetic field
generation found in simulations with shear.
For this purpose
a set of evolution equations for the response to imposed test fields is solved
with turbulent and mean motions calculated from the momentum and continuity equations.
The corresponding results for the electromotive force are used to
calculate turbulent transport coefficients.
The diagonal components of the turbulent magnetic diffusivity tensor
are found to be very close together,
but their values increase slightly with increasing
shear and decrease with increasing rotation rate.
In the presence of shear, the sign of the two off-diagonal components of
the turbulent magnetic diffusion tensor is the same
and opposite to the sign of the shear. This implies that
dynamo action from the shear--current effect is impossible,
except perhaps for high magnetic Reynolds numbers.
However, even though there is no alpha effect on the average,
the components of the $\alpha$ tensor display Gaussian fluctuations around
zero. These fluctuations are strong enough to drive an incoherent alpha--shear dynamo.
The incoherent shear--current effect, on the other hand, is found to be subdominant.
\end{abstract}


\section{Introduction}

Many of the stellar and planetary
magnetic fields are believed to be the result of a dynamo process
that converts kinetic energy from turbulent motions and shear into
magnetic energy.
A particular challenge consists in explaining the field on length scales that
exceed the scale of the turbulence.
This topic has traditionally been addressed within
the framework of mean--field electrodynamics (Krause \& R\"adler 1980).

Over the decades the applicability of this theory has repeatedly been
questioned (e.g., Piddington 1981, Vainshtein \& Cattaneo 1992).
Meanwhile, direct simulations of hydromagnetic turbulence have begun
to show dynamo action (Meneguzzi et al.\ 1981, Meneguzzi \& Pouquet 1989,
Nordlund et al.\ 1992, Brandenburg et al.\ 1996, Cattaneo 1999).
In some particular cases, large-scale fields are being generated
(Glatzmaier \& Roberts 1995, Brandenburg et al.\ 1995, Brandenburg 2001)
which raises the question about the mechanism responsible for this phenomenon.
In cases where the flow is systematically non-mirror symmetric
the association with an $\alpha$ effect is obvious.
However, there are now also examples of nonhelical large-scale dynamos
owing to turbulence under the influence of shear alone (Brandenburg 2005a,
Yousef et al.\ 2007).
Their interpretation is not straightforward, because several
possible mechanisms have been proposed that might produce dynamo action
from turbulence and shear alone, i.e.\ {\it without} rotation and stratification
that otherwise would have been the main ingredients of an $\alpha$ effect.
The most detailed investigations have been carried out in connection with the
so-called shear--current effect (Rogachevskii \& Kleeorin 2003, 2004,
R\"adler \& Stepanov 2006, R\"udiger \& Kitchatinov 2006).
Another possibility is a magnetic $\alpha$ effect that is driven by
a current helicity flux, as was suggested by Vishniac \& Cho (2001;
see also Brandenburg \& Subramanian 2005c).
A third possibility might be an incoherent (random) $\alpha$ effect
with zero mean and finite variance, suggested by Vishniac \& Brandenburg (1997)
in connection with accretion discs (see also Sokolov 1997, Silant'ev 2000,
Fedotov et al.\ 2006, Proctor 2007).
The only reliable way to determine what is the dominant effect is to calculate
all relevant components of the $\alpha$ and turbulent magnetic diffusivity tensors
in a general expansion of the electromotive force in terms of the
mean magnetic field.

The case considered in Brandenburg (2005a) is unnecessarily complicated
because the shear employed there depends on two Cartesian coordinates.
A simpler possibility is to consider a shear flow
depending linearly on only one coordinate and we shall
pursue this idea in the present paper.
The shear--current effect and the incoherent $\alpha$ effect could then
still operate.
Because we will use periodic boundary conditions
there can be no magnetic helicity flux, so the
Vishniac \& Cho (2001) effect is then ruled out,
even though it could still, at least in principle, explain the
generation of a mean magnetic field in the simulations of Brandenburg (2005a),
which do possess a helicity flux.

In this paper we calculate all relevant components of $\alpha_{ij}$
and $\eta_{ijk}$ using the so-called test field method.
This method was introduced by Schrinner et al.\ (2005, 2007) in connection
with convection in a spherical shell and used later by
Brandenburg (2005b), Sur et al.\ (2007) and Brandenburg et al.\ (2008)
in connection with forced turbulence in Cartesian boxes.
The essence of this method consists in solving evolution equations
for the fluctuations of the magnetic field around suitably defined
test fields such that all relevant coefficients can be computed.

\section{Basic assumptions and method}

In the following we introduce first the mean electromotive force
and its relation to the mean magnetic field.
We then discuss the equations describing the turbulent flow that eventually
leads to this electromotive force and explain the test field method
used to calculate the coefficients which relate it to the mean field.
Particular attention is paid to the possibility that the shear--current effect
may lead to self--excitation of mean magnetic fields.

\subsection{The turbulent electromotive force}
\label{turbemf}

In mean--field electrodynamics the behavior of the mean magnetic field $\meanBB$
depends crucially on the mean electromotive force $\meanEMF = \overline{\uu \times \bb}$,
where $\uu$ and $\bb$ denote the deviations of the fluid velocity $\UU$
and the magnetic field $\BB$ from their mean parts
$\meanUU$ and $\meanBB$, respectively.
For sufficiently weak variations of $\meanBB$ in space and time,
and if there is no small--scale dynamo
producing a mean electromotive force on its own, we have
\EQ
\meanemf_i = \alpha_{ij} \meanB_j + \eta_{ijk} \partial
\meanB_j / \partial x_k
\label{eq01}
\EN
with tensors $\alpha_{ij}$ and $\eta_{ijk}$ determined by $\uu$ and $\meanUU$.

In this section it is sufficient to define mean quantities like $\meanEMF$ or $\meanBB$,
referring to Cartesian coordinates $(x, y, z)$,
simply by averaging over all $x$ and $y$.
Below, in \Sec{testfield} a different definition will be introduced
that covers and refines this simple one.
Clearly $\meanBB$ can now no longer depend on $x$ and $y$ and hence all
its first--order derivatives can be expressed by the components
of $\nab \times \meanBB = ( - \partial \meanB_y /\partial z, \partial \meanB_x /\partial z, 0)$.
Slightly deviating from the usual notation, in which $\nab \times \BB$
is equal to $\mu_0 \JJ$ where $\mu_0$ is the vacuum permeability,
we put in this paper simply $\nab \times \BB = \JJ$,
being aware that $\JJ$ is then no longer exactly the electric current density.
Instead of (\ref{eq01}) we may now write
\EQ
\meanemf_i = \alpha_{ij} \meanB_j - \eta_{ij} \meanJ_j
\label{eq03}
\EN
with a new tensor $\eta_{ij}$ defined such that $\eta_{i1} = \eta_{i23}$ and $\eta_{i2} = - \eta_{i13}$.
As $\meanJ_3 = 0$ the $\eta_{i3}$ are without interest and we may put them equal to zero.

We further consider the background turbulence, which occurs in the
absence of rotation or shear, as homogeneous, isotropic and
mirror--symmetric.
Then we have even under the influence of rotation or
shear $\alpha_{ij} = 0$; see \App{JustificationAlp0}.

As for $\eta_{ij}$ consider first the case of rotation of the
fluid with an angular velocity $\OO$, which defines the Coriolis and centrifugal forces
and is assumed to be aligned with the $z$--axis.

The actual turbulence is then again homogeneous but no longer isotropic.
Instead it is axisymmetric with respect to the $z$--axis,
that is, all mean quantities depending on the turbulent velocity field are invariant
under arbitrary rotations about the $z$--axis.
We may then conclude by usual symmetry arguments that
\EQ
\eta_{ij} = \eta_0 \delta_{ij} + \delta \epsilon_{ijk} \hat{\Omega}_k
    + \delta' \hat{\Omega}_i \hat{\Omega}_j \, ,
\label{eq04}
\EN
where $\hat{\OO} = \OO / \Omega$ with $\Omega = |\OO|$,
and $\eta_0$, $\delta$ as well as
$\delta'$ are spatially constant coefficients, which may depend on $\Omega$.
So we arrive at
\EQ
\meanEMF = - \eta_0 \meanJJ + \delta \hat{\OO} \times \meanJJ \, .
\label{eq05}
\EN
Since $\meanJ_3 = 0$ the $\delta'$ term in \eq{eq04} is without influence.
The last term in \Eq{eq05} describes the $\OO \times \meanJJ$ effect (R\"adler 1969).
Whereas $\eta_0$ approaches a nonzero value as $\Omega \to 0$
(the value determined by the background turbulence),
$\delta$ vanishes like $\Omega$.
Note that $\meanemf_z = 0$.

Consider next the case with shear defined by the velocity $\UU^S = (0, Sx, 0)$.
Now the actual turbulence is again homogeneous but no longer axisymmetric.
In view of the application of symmetry arguments we consider $\UU^S$ first
in the more general (coordinate--independent) form $S \gggg\, (\hh \cdot \xx)$
where $\gggg$ and $\hh$ are unit vectors which are orthogonal to each other,
and $\xx$ is the position vector.
The only available construction elements for $\eta_{ij}$ are then $\delta_{kl}$, $\epsilon_{klm}$,
$\gggg$ and $\hh$,
for due to the
homogeneity of the turbulence, $\eta_{ij}$ cannot depend on $\xx$.
Thus we have
\EQ
\eta_{ij} = \eta_0 \delta_{ij} + \kappa_1 g_i g_j + \kappa_2 h_i h_j
    + \kappa_3 g_i h_j + \kappa_4 g_j h_i + \cdots
\label{eq06}
\EN
where $\eta_0$, $\kappa_1$, $\kappa_2$, $\kappa_3$ and $\kappa_4$ are spatially constant coefficients
and the dots stand for additional terms containing $\epsilon_{klm}$.
The aforementioned coefficients may depend on $S$.
(A dependence on scalars defined by $\gggg$ and $\hh$ is without interest since $\gggg^2 = \hh^2 = 1$
and $\gggg\cdot \hh = 0$.)
The terms containing $\epsilon_{klm}$ have structures like $\epsilon_{ijk} g_k$
or $g_i \epsilon_{jkl} g_k h_l$.
Since $S \gggg \, (\hh \cdot \xx)$ is invariant under simultaneous sign changes of $\gggg$ and $\hh$,
$\eta_{ij}$ must also have this property and so these terms have to be cancelled.
Returning now to $\UU^S = (0, Sx, 0)$, that is
$\gggg = (0, 1, 0)$ and $\hh = (1, 0, 0)$,
we see that
\EQ
\meanEMF = - \eta_0 \meanJJ - \kkappa \cdot \meanJJ
\label{eq07}
\EN
with
\EQ
\kkappa = \pmatrix{\kappa_{11} & \kappa_{12} & 0
\cr \kappa_{21} & \kappa_{22} & 0 \cr 0 & 0 & 0} \, .
\label{eq09}
\EN
This covers the ``shear--current effect'' (Rogachevskii \& Kleeorin 2003).
We may assume that $\eta_0$ is independent of $S$
(that is, it is determined by the background turbulence alone).
Then $\kappa_{11}$ and $\kappa_{22}$ are even functions of $S$ that vanish like $S^2$ as $S \to 0$,
whereas $\kappa_{12}$ and $\kappa_{21}$ are odd functions that vanish like $S$.
Again we have $\meanemf_z = 0$.

In both cases, with rotation or with shear, we may restrict our attention to
\EQ
\meanemf_i = -\eta_{ij} \meanJ_j \, , \quad 1\leq i,j \leq 2 \, .
\label{eq10}
\EN
The four quantities $\eta_{ij}$ are simply related to
$\eta_0$ and $\delta$, or $\eta_0$ and the $\kappa_{ij}$ respectively.

\subsection{Turbulence with rotation or shear}
\label{turbulence}
We consider a compressible fluid
satisfying an isothermal equation of state.
In the absence of rotation or shear the momentum and continuity equations
can be written in the form
\EQ
{\partial\UU\over\partial t}=-\UU\cdot\nab\UU-c_{\rm s}^2\nab\ln\rho +\ff+\FF_{\rm visc},
\label{dUU}
\EN
\EQ
{\partial\ln\rho\over\partial t}=-\UU\cdot\nab\ln\rho-\nab\cdot\UU,
\label{dlnrho}
\EN
where $c_{\rm s}$ is the sound speed, here considered as constant,
$\rho$ the mass density and $\ff$ a random forcing function.
Furthermore, $\FF_{\rm visc}=\rho^{-1}\nab\cdot2\rho\nu\SSSS$ is the viscous force, and
${\sf S}_{ij}={1\over2}(U_{i,j}+U_{j,i})-{1\over3}\delta_{ij}\nab\cdot\UU$
is the traceless rate of strain tensor.

To come as close as possible to the assumptions on the background turbulence adopted above,
i.e., homogeneity, isotropy and mirror--symmetry,
the forcing function $\ff$
was specified
for a cubic domain of size $L\times L\times L$
as follows.
During each time--step $\ff$ is a single transverse (solenoidal) plane wave
proportional to $\kk_{\rm f}\times\eee$ where the wavevector $\kk_{\rm f}$
is taken randomly from a set of pre-defined vectors with
components being integer multiples of $2\pi/L$
and moduli in a certain interval around an average value
which we simply denote by $k_{\rm f}$,
and $\eee$ is an arbitrary random unit vector not aligned with $\kk_{\rm f}$.
The corresponding scale, $2\pi/k_{\rm f}$, is also referred to as
the energy-carrying scale of the turbulence.
Moreover, the time dependence of $\ff$ is designed to mimic
$\delta$--correlation, which is a simple and commonly used
form of random driving (cf.\ Brandenburg 2001).
Nevertheless, owing to inertia,
the correlation time of the turbulent velocity is of course finite,
even for perfect $\delta$--correlation of the forcing.

As mentioned above, we simulate the turbulence in a finite domain
using (shearing--) periodic boundary conditions (\Sec{simul}).
Then the background turbulence can be at no instant
in a strict sense homogeneous, isotropic or mirror-symmetric.
It would approach these properties if the ratio of the size of this domain
and the scale of the forcing function (that is $k_{\rm f}/k_1$ with $k_1=2\pi/L$)
became very large.
There are, however, practical bounds on this ratio.
For moderate values, which we have to accept, the background turbulence approaches
the mentioned properties only after averaging over long times.
Then, of course, the turbulence appears also as statistically steady.
By these reasons mean quantities, that is, averages over $x$ and $y$,
which are derived from the turbulence, show still fluctuations in $z$ and $t$,
and these disappear after averaging over sufficiently long time intervals.

When rotation is added, two new terms arise on the right hand side
of \Eq{dUU}, the Coriolis force, $-2\OO\times\UU$, and the
centrifugal force, $(\OO\times\xx)\times\OO$.
The latter is unimportant for weak compressibility,
to which we restrict ourselves in the following,
and this term would also not be compatible with periodic boundary conditions,
so it is neglected.

Turning now to the case with shear we redefine the velocity $\UU$
by splitting off the shear term $\UU^S$, that is $\UU
\to \UU+\UU^S$. This implies \EQ \UU\cdot\nab\UU\rightarrow
\UU\cdot\nab\UU+\UU^S\cdot\nab\UU+\UU\cdot\nab\UU^S+\UU^S\cdot\nab\UU^S.
\label{NonlinearTerm} \EN The second term on the right hand side
corresponds to an additional advection with the mean flow and will
be subsumed in the definition of an advective derivative, \EQ
\DDD/\DDD t\equiv\partial/\partial t+Sx\;\partial/\partial y. \EN
The third term is equal to $SU_x\yyy$, where $\yyy$ is the unit vector in the $y$ direction.
The last term in
\Eq{NonlinearTerm} vanishes. Thus, equations (\ref{dUU}) and
(\ref{dlnrho}) turn into \EQ {\DDD\UU\over\DDD
t}=-\UU\cdot\nab\UU+SU_x\yyy -c_{\rm s}^2\nab\ln\rho+\ff+\FF_{\rm
visc}, \label{dUUS} \EN and \EQ {\DDD\ln\rho\over\DDD
t}=-\UU\cdot\nab\ln\rho-\nab\cdot\UU. \label{dlnrhoS}\EN

It should be noted that $\UU$ resulting from these equations is not purely turbulent,
but also contains a large-scale flow which provides an additional shear and therefore a mean vorticity.
This is qualitatively suggestive of a hydrodynamic mean--field effect analogous to the shear--current
effect; see Elperin et al.\ (2003).

In this paper we deal, apart from one exception, with the purely kinematic problem,
so there is no Lorentz force in \Eq{dUUS}.
In \Sec{LSfields} the fully nonlinear problem is considered and hence
the Lorentz force is included in the momentum equation.

\subsection{Test field method}
\label{testfield}

Proceeding now to consequences of the induction equation we consider primarily the case of shear,
in which the fluid velocity is $\UU+\UU^S$.
In the case of rotation we have to put $\UU^S$ equal to zero.
We further represent $\BB$ according to $\BB = \nab \times \AAA$ by a vector potential $\AAA$.
Uncurling the induction equation and using a suitable gauge transformation of $\AAA$ we find%
\footnote{Note that $\UU^S\times\BB$ can be written as
$(\UU^S\times\nab\times\AAA)_i=U^S_j(A_{j,i}-A_{i,j})$. The second
term is an advection term and the first term can be written as
$U^S_j A_{j,i}=-U^S_{j,i} A_j$ plus a gradient term that can be
removed with a gauge transformation. Note also that $U^S_{j,i}
A_j=SA_y\hat{x}_i$.}
\EQ
{\DDD\AAA\over\DDD t}=-SA_y\xxx+\UU\times\BB-\eta\JJ \, .
\label{dAdt}
\EN
This equation as well as those derived from it in what follows
apply to the case of rotation if ${\DDD/\DDD t}$ is replaced by $\partial / \partial t$
and $S$ is put equal to zero.

Now we define a mean field $\overline{F}$ belonging to the field $F$ as
\EQ
\overline{F} (x, y, z) = \frac{1}{L^2} \int_{-L/2}^{L/2} \int_{-L/2}^{L/2}
    F(x + \xi, y + \eta, z) \, \dd \xi \, \dd \eta
\label{average}
\EN
with $L$ as specified above. The following comments on the definition in \Eq{average} as well as equations
\eq{meanfield} and \eq{eq11} below apply, however, even if $L$ is an arbitrary length, not necessarily
related to the domain size.
Our definition \eq{average} implies that averaging of $F$ commutes with taking any derivatives of $F$
with respect to $x$, $y$, $z$ or $t$, that is, the sequence of these operations can be changed.
In what follows we also use the rule $\overline{\overline{F} G} = \overline{F} \, \overline{G}$,
which applies exactly if $\overline{F}$ is independent of $x$ and $y$,
and has otherwise to be considered as an approximation.
It is however only needed in cases in which that independence of $x$ and $y$ can be justified,
that is, in which it applies exactly.
Clearly,
$\overline{F}$ is independent of $x$ and $y$ if $F$ is periodic in $x$ and $y$
with the period length $L$.
We note further that, owing to (\ref{average}), we have $\overline{x} = x$,
and that therefore $\UU^S$ has to be considered as a mean field.

Taking now the average of (\ref{dAdt}) we obtain
\EQ
{\DDD\meanAA\over\DDD t}
=-S\meanA_y\xxx+\meanUU\times\meanBB+\overline{\uu\times\bb}-\eta\meanJJ.
\label{meanfield}
\EN
In view of the determination of $\meanEMF = \overline{\uu \times \bb}$
we are interested in $\bb = \nab \times \aaa$, where  $\aaa=\AAA-\meanAA$.
Taking the difference between equations (\ref{dAdt}) and (\ref{meanfield})
we obtain
\EQ
{\DDD\aaa\over\DDD t}=-Sa_y\xxx+\meanUU\times\bb+\uu\times\meanBB
    +\uu\times\bb-\overline{\uu\times\bb}-\eta\jj \, ,
\label{eq11}
\EN
where $\jj = \JJ - \meanJJ$.

In order to determine the quantities $\eta_{ij}$ introduced above
we specify $\meanBB$ in the relevant relations such that it is equal to one of the elements
out of a set of test fields, $\meanBB^{q}$,
and denote the corresponding $\meanEMF$, $\meanJJ$, etc., by $\meanEMF^{q}$, $\meanJJ^{q}$, respectively.
Then, in particular, \Eq{eq10} turns into
\EQ
\meanemf^{q}_i = - \eta_{ij} \meanJ^{q}_j \, , \quad 1\leq i,j \leq 2 \, .
\label{eq13}
\EN
After having calculated the $\meanEMF^{q}$ numerically for two properly chosen $\meanBB^{q}$
we may then determine the four $\eta_{ij}$.

For the calculation of the $\meanEMF^{q}$ we apply (\ref{eq11})
\footnote{In the corresponding eq. (27) of Brandenburg
(2005b), the $\meanUU$ term is incorrect. However this did not
affect his results because $\meanUU$ either vanished or it
consisted only of a shearing motion that was treated correctly in the
code.},
\EQ
{\DDD\aaa^{q}\over\DDD t}= -Sa_y^{q}\xxx+\meanUU\times\bb^q+\uu\times\meanBB^{q}
+\uu\times\bb^{q}-\overline{\uu\times\bb^{q}} -\eta\jj^{q}.
\label{daapq}
\EN
Although an additional mean flow can develop in some of the simulations with
shear (see above), this term is still weak and is neglected in the following,
hence we put $\meanUU={\bf0}$ in \Eq{daapq}.

As test fields $\meanBB^{q}$ we may use, e.g., the fields $\meanBB^{q {\rm c}}$ defined by
\EQ
\meanBB^{1 {\rm c}} =B (\cos kz, 0, 0)  \, , \quad
    \meanBB^{2 {\rm c}} =B (0, \cos kz, 0)
\label{eq15}
\EN
with a constant $B$ and a constant wavenumber $k$.
Denoting the corresponding $\meanEMF^{q}$ by $\meanEMF^{q {\rm c}}$ we find
\EQ
\meanemf^{1 {\rm c}}_i =  \eta_{i2} B k \sin kz \, , \quad
\meanemf^{2 {\rm c}}_i = -\eta_{i1} B k \sin kz \, , \quad i = 1, 2 \, .
\label{eq17}
\EN
After having calculated the $\meanEMF^{q {\rm c}}$ these equations allow us
to determine the $\eta_{ij}$.
In order to avoid difficulties at the zeros of $\sin kz$
it is useful to carry out the calculations with test fields $\meanBB^{q {\rm s}}$,
defined analogously to the $\meanBB^{q {\rm c}}$ but with $\sin kz$ instead of $\cos kz$.
For the corresponding $\meanEMF^{q {\rm s}}$ we find then equations analogous to (\ref{eq17})
but with $- \cos kz$ instead of $\sin kz$.
From this and equation~(\ref{eq17}) we obtain immediately
\begin{eqnarray}
\eta_{i1} &=& - (Bk)^{-1}\big( \meanemf^{2 {\rm c}}_i \sin kz - \meanemf^{2 {\rm s}}_i \cos kz \big)
\nonumber\\
\eta_{i2} &=& (Bk)^{-1}\big( \meanemf^{1 {\rm c}}_i \sin kz - \meanemf^{1 {\rm s}}_i \cos kz \big) \, , \quad
    i = 1, 2
\label{eq19}
\end{eqnarray}
(see also Brandenburg 2005b).

We recall that for homogeneous turbulence, which is considered here,
the $\eta_{ij}$ have to be independent of $z$.
That is, the $\cos kz$ and $\sin kz$ in (\ref{eq17}) and (\ref{eq19})
should be compensated by $z$--dependencies of the $\meanemf^{q {\rm c}}_i$ and $\meanemf^{q {\rm s}}_i$.
However, due to fluctuations (cf.\ \Sec{turbulence}), no perfect compensation can be expected.

We further recall that in (\ref{eq01}) and so also in (\ref{eq03})
all derivatives of $\meanBB$ that are higher than first--order have been ignored.
By this reason the results for the $\eta_{ij}$ obtained with the above test fields apply exactly
only in the limit $k \to 0$.
In general there is a dependence of the $\eta_{ij}$ on $k$.
This corresponds to a non--local connection
between $\meanEMF$ and $\meanBB$,
which is considered here only in a very weak sense (by taking into account first--order derivatives of $\meanBB$).
In a more general sense it is investigated in Brandenburg et al.\ (2008).
Here we have used $k = k_1$ where $k_1$ means the smallest finite wavenumber in the $z$-direction
in the domain in which the turbulence is simulated, $k_1 = 2 \pi / L$; see \Sec{simul}.

The results for $\eta_{ij}$ are also independent of the value of $B$.
If one wanted to address the question of nonlinearity, which is not the
purpose of this paper, one must also solve \Eq{dAdt}
and allow the resulting magnetic field
to feed back onto the flow via the Lorentz force.

For the discussion of the results concerning $\eta_{ij}$ we introduce the
quantities
\EQ
\eta_{\rm t} = \half (\eta_{11} + \eta_{22}) \, , \quad
    \eta_{\rm T} = \eta + \eta_{\rm t} \, , \quad
    \epsilon = \half (\eta_{11} - \eta_{22}) \, .
\label{eq21}
\EN
In the case of rotation we put further
\EQ
\delta = \half (\eta_{12} - \eta_{21})
\label{eq23}
\EN
and expect $\epsilon$ to be equal to zero,
while $\eta_{12}$ and $\eta_{21}$ have to have the same nonzero moduli, but opposite signs
so that $\delta$ is nonzero.
With shear, however, $\epsilon$ can be in general nonzero
and there is no simple relation between $\eta_{12}$ and $\eta_{21}$.

It is convenient to present $\eta_{\rm t}$ in normalized form and express
it in terms of the quantity
\EQ
\eta_{\rm t0}=\onethird u_{\rm rms}k_{\rm f}^{-1},
\EN
which corresponds to the result for $\eta_{\rm t}$ obtained
under the first order smoothing approximation
applied to the high conductivity limit under the assumption that the
correlation time is given by $(u_{\rm rms}k_{\rm f})^{-1}$,
i.e., that the Strouhal number is unity (cf.\ Brandenburg \& Subramanian 2005b).

\subsection{Dispersion relation}

In the case of rotation without shear there are only decaying solutions
of the mean--field equations.
This can be easily seen from the energy balance equation for the mean magnetíc field
(see R\"adler 1980).
The situation with shear alone is however different and
the possibility of a so--called shear--current dynamo is still under debate
(see Rogachevskii \& Kleeorin  2003, 2004, Brandenburg 2005b,
R\"adler \& Stepanov 2006, R\"udiger \& Kitchatinov 2006).
We look therefore for solutions of (\ref{meanfield})
with $\overline{\uu \times \bb}$ specified in the sense of (\ref{eq10}).
Using the ansatz  $\meanAA=\tilde{\AAA}\exp(\lambda t+\ii kz)$ with a generally complex $\lambda$
and any real $k$ satisfying $k\ll k_{\rm f}$ we find first
\EQ
\pmatrix{\lambda+(\eta_{\rm T}+\epsilon)k^2 &\eta_{12}k^2+S\cr
\eta_{21}k^2 & \lambda+(\eta_{\rm T}-\epsilon)k^2}
\pmatrix{\tilde{A}_{x}\cr\tilde{A}_{y}}={\bf0} \, .
\label{lamp}
\EN
The requirement of non--vanishing $\tilde{\AAA}$ poses an eigenvalue problem
for $\lambda$.
 The two eigenvalues, normalized to $\eta_{\rm T}k^2$, are
\EQ
{\lambda_\pm\over\eta_{\rm T}k^2}=-1\pm
\frac{1}{\eta_{\rm T}}\sqrt{(S / k^2 + \eta_{12}) \eta_{21} + \epsilon^2 } \, .
\label{lampm}
\EN
A necessary and sufficient condition for an exponentially growing solution
is that the radicand in (\ref{lampm}) is positive and that it exceeds $\eta^2_{\rm T}$.
If the $S$ term
dominates and the others are neglected this condition
turns into
\EQ
D_{\eta S}\equiv{S\eta_{21}\over(\eta_{\rm T} k)^2} > 1. \label{dyncond}
\EN
As $k$ can be made arbitrarily small (by making the domain size large enough)
this condition is always satisfiable if only  $S \eta_{21}>0$.
The neglect of the terms without $S$ in the radicand is justified if
$|\eta_{12} \eta_{21} + \epsilon^2| / \eta_{\rm T}^2 \ll |D_{\eta S}|$,
which can again always be guaranteed by sufficiently small $k$.
Under this condition the maximum growth rate with respect to $k$
is $S\eta_{21}/4\eta_{\rm T}$ and occurs at $k =
\sqrt{S\eta_{21}} / 2\eta_{\rm T}$. Consequently, as long as
$\eta_{21}$ can be considered linear in $S$ the maximum growth
rate is proportional to $S^2$
and the corresponding $k$ proportional to $S$.

\subsection{Simulations}
\label{simul}

For the numerical simulations
we use the \textsc{Pencil Code}\footnote{
\url{http://www.nordita.org/software/pencil-code}}, where
the test field algorithm has already been implemented.
We employ periodic boundary conditions in the $y$ and $z$ directions
and shearing--periodic boundary conditions in the $x$ direction
(Wisdom \& Tremaine 1988, Hawley et al.\ 1995)
and use a resolution of up to $256^3$ meshpoints
for the runs with the largest Reynolds numbers.
As mentioned above, a computational domain of size $L^3$ is used,
so the smallest finite wavenumber is $k_1=2\pi/L$.
As initial conditions for the hydrodynamic part
we assume vanishing velocity, $\UU={\bf 0}$,
and uniform density equal to some value $\rho_0$.
The initial condition in the test field calculations is $\aaa^q={\bf 0}$.
Owing to the use of periodic boundary conditions,
the total mass in the computational domain is conserved,
and therefore the mean density will be always equal to the initial value, $\bra{\rho}=\rho_0$,
where $\bra{\ldots}$ denotes a volume average.

In all investigations reported in this paper only weakly compressible turbulence
has been considered.
The Mach number
$u_{\rm rms}/c_{\rm s}$ did not exceed a value of the order of 0.1.

\section{Results for the diffusivity coefficients}
\label{Results}

As explained above,
the test field procedure yields the coefficients $\eta_{ij}$ first as functions of $z$ and $t$.
However, after averaging over sufficiently long time intervals,
we expect to approach the
results for homogeneous, isotropic, mirror--symmetric and statistically steady background turbulence,
in particular coefficients $\eta_{ij}$,
being independent of $z$ and $t$.
We present here results for the $\eta_{ij}$ gained by averaging of the `raw' data first over $z$
and then over time.
In this context the effect of averaging over $z$ consists in a first reduction of the temporal
fluctuations.
This appears plausible in the picture in which the domain contains
a finite number of turbulent ``eddies" (Hoyng 1993).
We may interpret them as different realizations of a specific eddy
and thus the average over the $x$, $y$ and $z$ of a given domain
as an average over the ensemble of these realizations.
When accepting the principle that the ensemble average is equivalent to a time average
we see that the effect of averaging the original $\eta_{ij}$ over $z$
is just equivalent to some temporal smoothing.
After having averaged over $z$, time averages are then taken over a suitable stretch
of the full time series where these averages are approximately steady.
We use the time series further to calculate error bars as the maximum departure
between these averages and the averages obtained from one of three equally long
subsections of the full time series.

Important control parameters that are being varied include
the hydrodynamic and magnetic Reynolds numbers, $\Rey$ and $\Rm$,
as well as the magnetic Prandtl number $\Pm$, with
\EQ
\Rey=u_{\rm rms}/(\nu k_{\rm f}),\quad
\Rm=u_{\rm rms}/(\eta k_{\rm f}),\quad
\Pm=\nu/\eta \, .
\EN
In the case of rotation we define further the Coriolis number $\Co$
and in the case of shear we define the parameter $\mbox{Sh}$,
\EQ
\Co=2\Omega/(u_{\rm rms}k_{\rm f}),\quad
\mbox{Sh}=S/(u_{\rm rms}k_{\rm f}) \, .
\EN
We note that $\Co$, like $\Omega$, is never negative.
In all cases with shear presented below, $S$ and thus $\mbox{Sh}$ are negative.
For most of the calculations we use $k_{\rm f}/k_1=5$,
except in \Sec{LSfields} where $k_{\rm f}/k_1=10$.
In both cases the range of forcing wavenumbers is $k_f \pm k_1/2$.

\subsection{Effect of rotation}
\label{EffectRot}

\begin{figure}[t!]
\centering\includegraphics[width=\columnwidth]{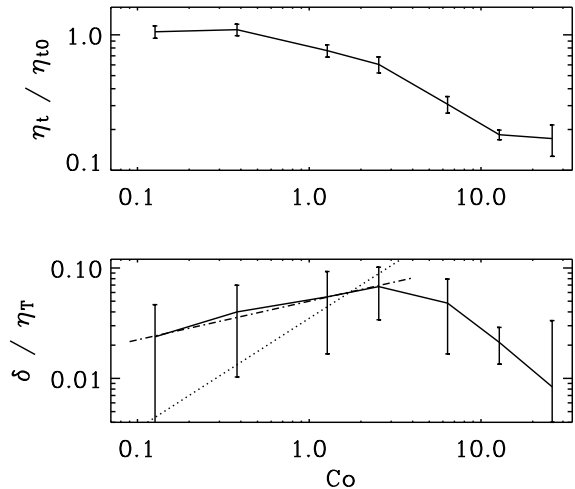}\caption{
Dependence of the normalized $\eta_{\rm t}$ and $\delta$ on $\Co$
for homogeneous turbulence with rotation for $\Rey=1.3$ and $\Rm=13$.
The vertical lines indicate error bars.
Note that there is a maximum of $\delta/\eta_{\rm T}$ at $\Co\approx3$.
For $\Co<3$ the results for $\delta/\eta_{\rm T}$ are best described by
$\delta/\eta_{\rm T}\approx0.05\times\Co^{0.35}$, given by the dash-dotted line,
but also a linear dependence, $\delta/\eta_{\rm T}\approx0.035\times\Co$,
indicated by the dotted line, is compatible within error bars.
}\label{kinrot_vs_Co}\end{figure}

\begin{figure}[t!]
\centering\includegraphics[width=\columnwidth]{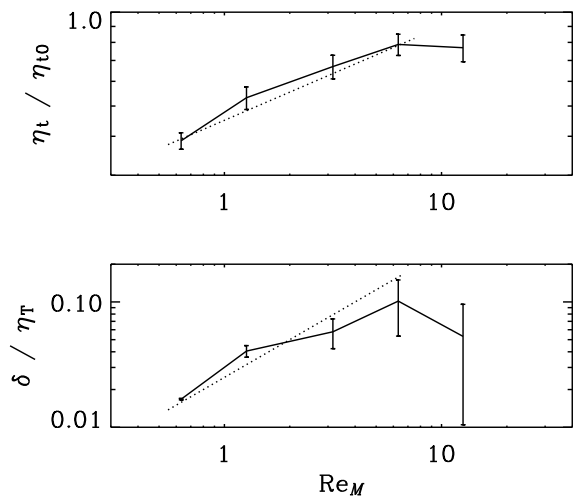}\caption{
Dependence of the normalized
$\eta_{\rm t}$ and $\delta$ on $\Rm$
for homogeneous turbulence with rotation for
$\Rey=1.3$ and $\Co=1.3$.
The dotted lines show the power law fits
$\eta_{\rm t}/\eta_{\rm t0} = 0.45\, \Rm^{0.3}$
and $\delta/\eta_{\rm T} = 0.025\, \Rm$ which apply for $\Rm < 7$.}
\label{kinrot_Re14}\end{figure}

In the case of rotation ($\Co \neq 0$), but without shear (\mbox{Sh}=0),
the coefficients $\eta_{\rm t}$ and $\delta$ are relevant.
\FFig{kinrot_vs_Co} shows their dependence on the Coriolis number $\Co$
for fixed Reynolds numbers, $\Rey=1.3$ and $\Rm=13$.
We see that $\eta_{\rm t}$ shows a drastic decline when $\Co$ approaches
and exceeds unity.
This can be understood as a consequence of an evolving Taylor-Proudman state
of the turbulent flow.
Clearly $\delta$ is positive.
The ratio $\delta / \eta_{\rm T}$ first increases with $\Co$,
but it begins to decline when $\Co$ has exceeded a value of about 3.
In \Fig{kinrot_Re14} the dependence of $\eta_{\rm t}$ and $\delta$ on $\Rm$ is given
for $\Rey=16$ and $\Co=1.3$.
As $\Rm$ is increased, $\eta_{\rm t}$ and $\delta$ increase for $\Rm<10$.

According to the considerations in \Sec{turbemf} we have to expect that the diagonal elements
$\eta_{11}$ and $\eta_{22}$ of the magnetic diffusivity tensor coincide.
Indeed, the observed values of $\epsilon$ (not shown) are only
of the order of the errors.

Our results are consistent with those obtained in the framework
of the second--order correlation approximation, see \App{SOCA}.
We take this consistency as a confirmation of the correctness of the test--field method.

\begin{figure}[t!]
\centering\includegraphics[width=\columnwidth]{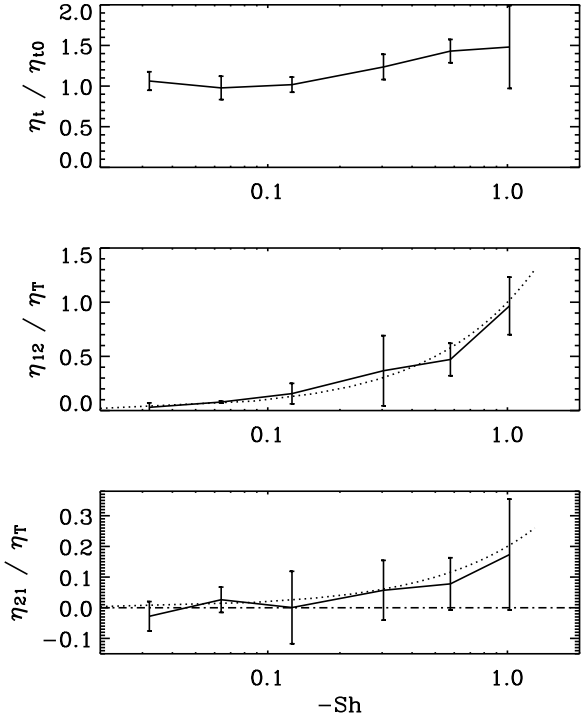}\caption{
Dependence of $\eta_{\rm t}$ (normalized by $\eta_{\rm t0}$)
as well as $\eta_{12}$ and $\eta_{21}$
(both normalized by $\eta_{\rm T}$) for homogeneous turbulence with shear on $\mbox{Sh}$
for $\Rey=1.4$ and $\Rm=14$.
The dotted lines represent linear dependencies on $\mbox{Sh}$.
}\label{kinshear_vs_shear}\end{figure}

\begin{figure}[t!]
\centering\includegraphics[width=\columnwidth]{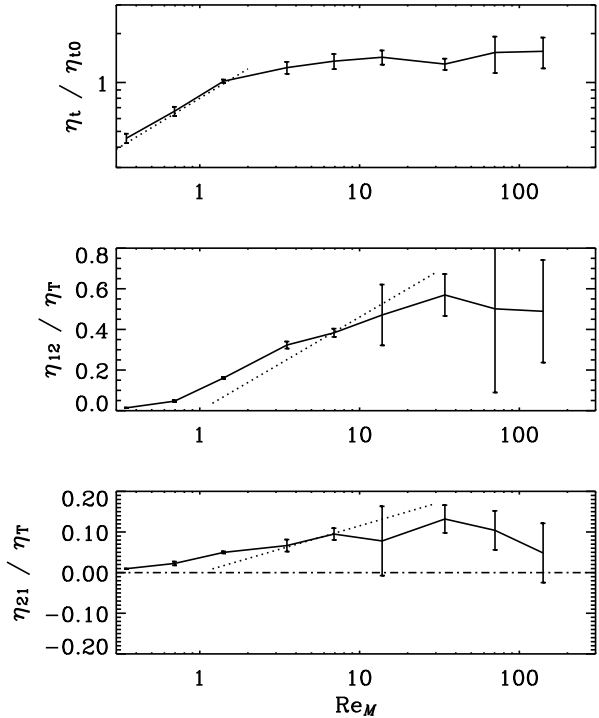}\caption{
Dependences of $\eta_{\rm t}$ (normalized by $\eta_{\rm t0}$),
as well as $\eta_{12}$ and $\eta_{21}$
(both normalized by $\eta_{\rm T}$) for homogeneous turbulence with shear on $\Rm$
for $\Rey=1.4$ and $\mbox{Sh}=-0.6$.
The dotted lines correspond to 
$\eta_{\rm t} / \eta_{\rm t0} = 0.8\Rm^{0.6}$,
$\eta_{12} / \eta_{\rm T} = 0.2\ln \Rm$
and $\eta_{21}/ \eta_{\rm T} = 0.05 \ln \Rm$
and illustrate that $\eta_{12}/\eta_{\rm T}$ and $\eta_{21}/\eta_{\rm T}$
vary only weakly with $\Rm$.
}\label{kinshear_Re14}\end{figure}

\subsection{Effect of shear}
\label{EffectShear}

We now discuss the case of shear ($\mbox{Sh}\neq0$) in the absence of rotation ($\Co = 0$).
\FFig{kinshear_vs_shear} demonstrates that the value of $\eta_{\rm t}/\eta_{\rm t0}$
clearly exceeds unity for not too small values of $|\mbox{Sh}|$,
that is, shear leads to a slight enhancement of the turbulent magnetic diffusivity.
At the same time, for negative values of $\mbox{Sh}$, both $\eta_{12}$ and $\eta_{21}$
attain finite positive values.
In \Figs{kinshear_Re14}{kinshear_Pm20} these quantities are shown as functions of $\Rm$,
with $\Rey=1.4$ and $\mbox{Sh}=-0.6$, or $\Pm=20$, respectively.

\begin{figure}[t!]
\centering\includegraphics[width=\columnwidth]{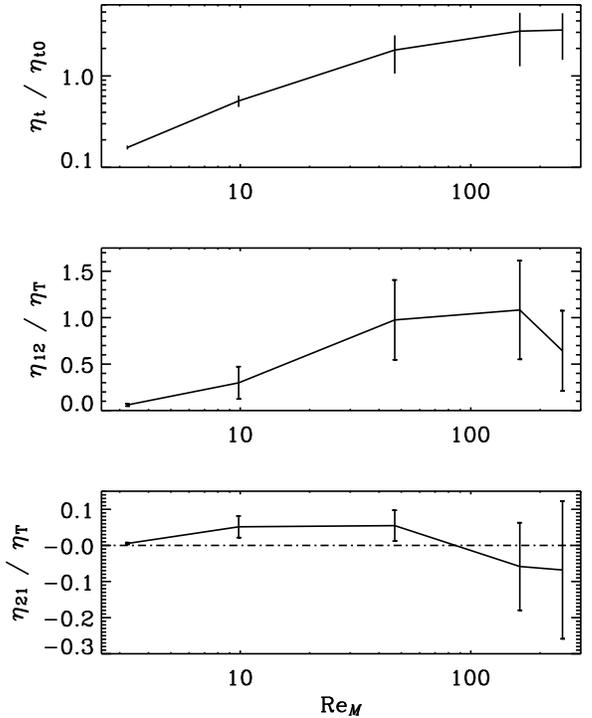}\caption{
Same as \Fig{kinshear_Re14}, but for a fixed magnetic Prandtl number, $\Pm=20$.
Here, $\Rey=\Rm/20$ is not constant. Because $u_{\rm rms}$
increases with increasing $\Rm$, $\mbox{Sh}$ is also not constant
and varies between $-2.5$ (for $\Rey=0.16$) and $-0.3$ (for $\Rey=13$).
Note that $\eta_{21}$ turns negative at about $\Rm=100$.
However, the errors are larger than the mean.
}\label{kinshear_Pm20}\end{figure}

\begin{figure}[t!]
\centering\includegraphics[width=.9\columnwidth]{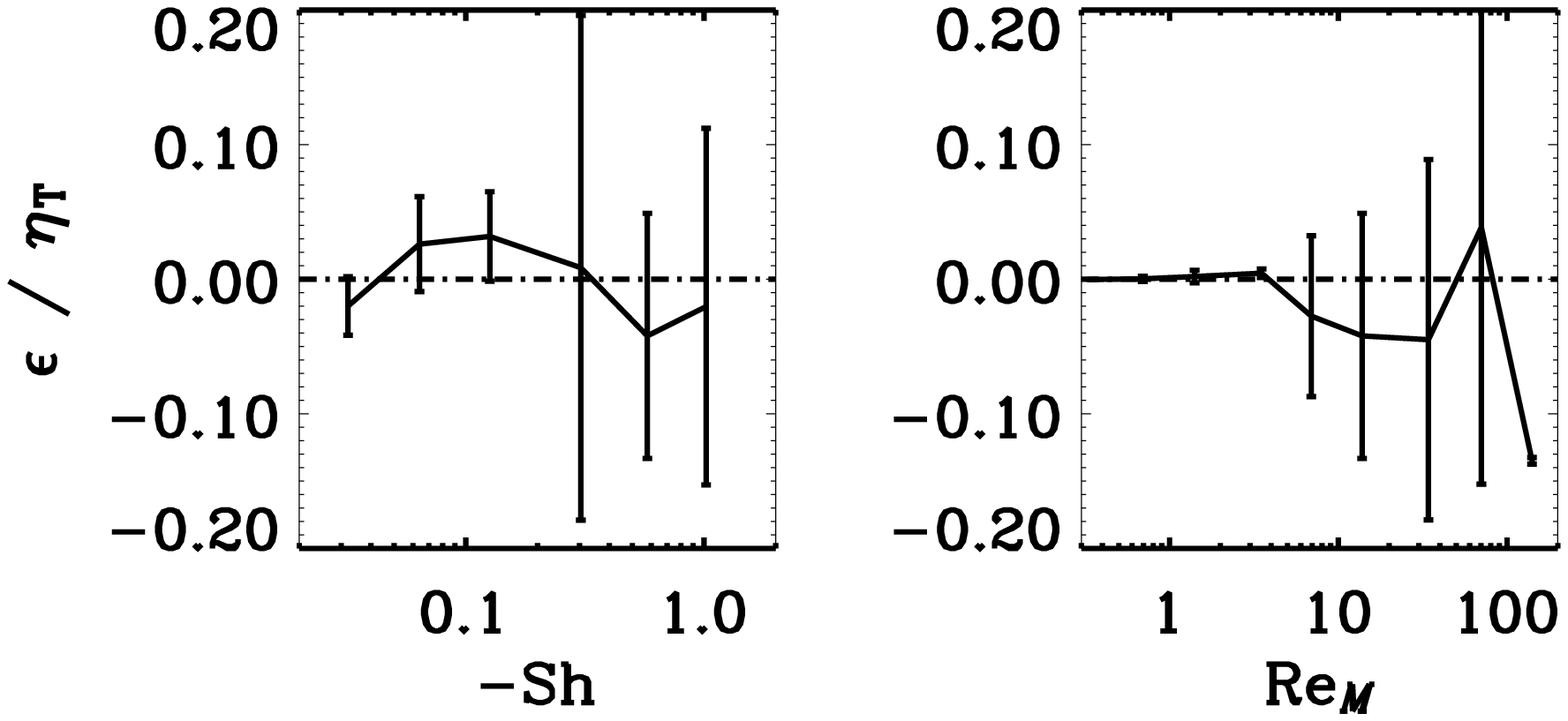}\caption{
Dependences of $\epsilon$ (normalized by $\eta_{\rm T}$) on $\mbox{Sh}$
({\it left}) and on $\Rm$ ({\it right})
for the same runs as in \Fig{kinshear_vs_shear} and \Fig{kinshear_Re14}, respectively.
}\label{kinshear_vs_shear_eps}\end{figure}

We recall the dynamo condition (\ref{dyncond}).
Since in all our simulations $S$ is negative, a dynamo would be possible for negative $\eta_{21}$ only.
In \Fig{kinshear_Pm20}, with $\Pm = 20$, we see indeed negative $\eta_{21}$ for high $\Rm$.
Considering the large error bars, however,
we may hardly conclude that a dynamo is really possible.
In general the errors could be reduced by extending the time series.
However, for large $\Rm$ small-scale dynamo action occurs that
introduces additional fluctuations whose amplitude increases exponentially
with time, and so we have to stop the calculation.
One remedy might be to reset $\bb^q$ in regular time intervals,
but this has not been done yet.

As the considerations of \Sec{turbemf} show, there is no general reason for an equality
of the two diagonal elements $\eta_{11}$ and $\eta_{22}$ of the diffusivity tensor,
that is, $\epsilon$ does not need to be equal to zero.
As shown in Fig.~\ref{kinshear_vs_shear_eps},
for $\Rm$ of order 10 and above, $\epsilon$ may deviate from zero, but its
value is of the order of the error.
Again, our numerical results are in agreement with results obtained
in the second--order correlation approximation, see \App{SOCA}.

\section{Comparison with direct simulations and simplified models}

\subsection{Large--scale fields in simulations with shear}
\label{LSfields}

\begin{figure}[t!]
\centering\includegraphics[width=\columnwidth]{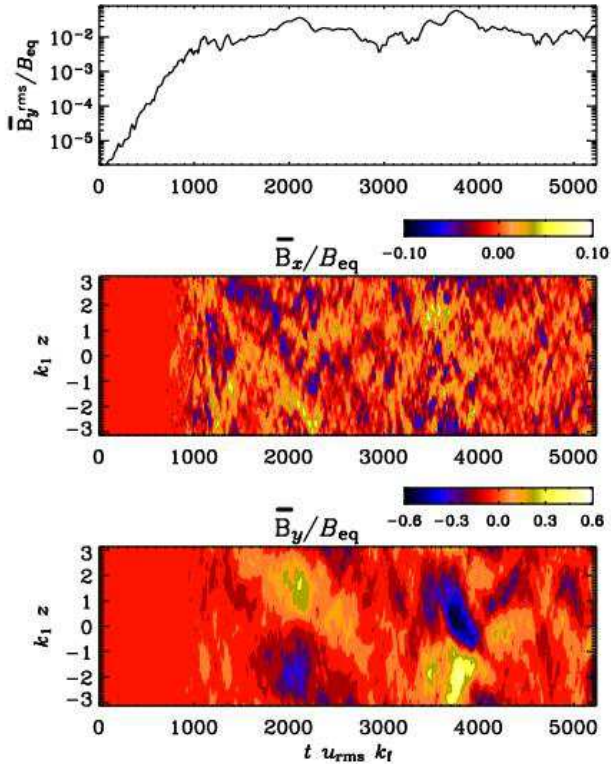}\caption{
Time dependence of the rms value (with respect to $z$)
of $\meanB_y$ ({\it top})
and space-time diagrams $\meanB_x(z,t)$ and $\meanB_y(z,t)$
(all in units of $B_{\rm eq}$, where
$B_{\rm eq} = \sqrt{\mu_0 \bra{\rho u^2}}$)
from a direct simulation with $\Rm=130$, $\Pm=7$, $k_{\rm f}/k_1=10$
and $\mbox{Sh}\approx-0.15$.
The top panel demonstrates the initial exponential growth of the
mean field (the growth rate is $0.009\,u_{\rm rms}k_{\rm f}$).
The other panels show episodes of large scales in $z$
especially in the $\meanB_y$ component.
}\label{pmeanfield_w256a}\end{figure}

\begin{figure}[t!]
\centering\includegraphics[width=\columnwidth]{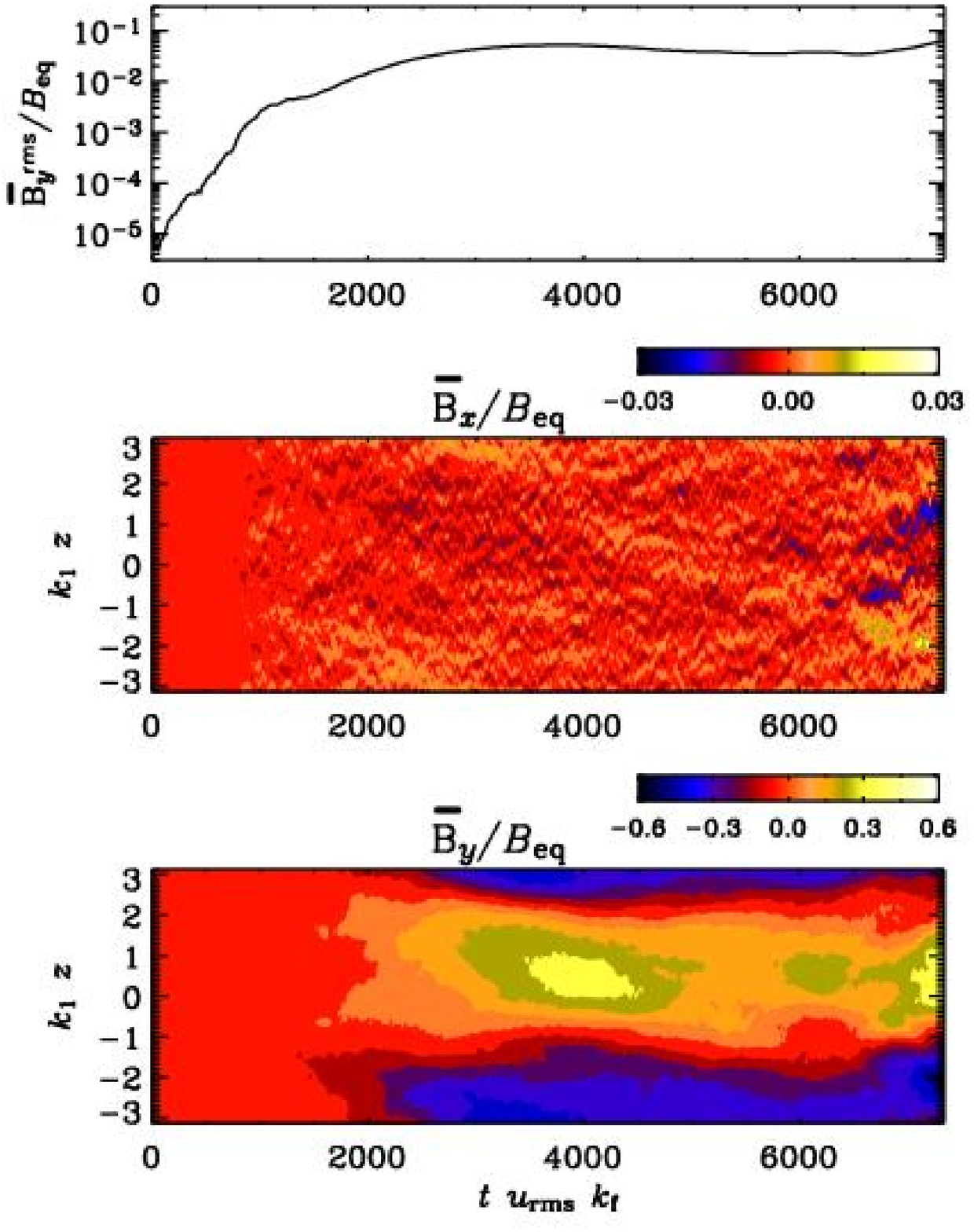}\caption{
Same as \Fig{pmeanfield_w256a}, but for $\Rm=200$ and $\Pm=20$.
Initially, the field grows exponentially at a rate $0.012\,u_{\rm rms}k_{\rm f}$.
Note that the mean field is nearly steady.
}\label{pmeanfield_w256b}\end{figure}

We report now on calculations with the original induction equation \eq{dAdt},
instead of the test field equations \eq{daapq},
together with the hydrodynamic equations (\ref{dUU}) and (\ref{dlnrho}).
In the momentum equation (\ref{dUU}), however, the Lorentz force was restored,
thus providing a nonlinear feedback of the magnetic field.
In all cases we used $\mbox{Sh}=-0.15$ and $k_f/k_1=10$.
The results are shown in \Figs{pmeanfield_w256a}{pmeanfield_w256b} for
two different combinations of $\Rm$ and $\Pm$.
In both cases there is an initial phase where the mean field grows
exponentially.
Mean fields with a particularly prominent $\meanB_y$ component occur.
The $\meanB_x$ component seems to be in antiphase with $\meanB_y$,
as expected for negative shear, but this component is much more noisy.
Furthermore, for $\Pm=7$ and $\Rm=130$ (\Fig{pmeanfield_w256a})
there are episodes where $\meanB_y$ fades away and later reemerges,
but possibly with the opposite orientation.
Similar results (not shown here)
have also been obtained for smaller values of $\Rm$.
In the case with $\Pm=20$ and $\Rm=200$ (\Fig{pmeanfield_w256b}), however,
$\meanB_y$ keeps the same orientation throughout the run.

\subsection{Magnitude and effect of fluctuations}

In \Sec{EffectShear} we have seen that the sign of $\eta_{21}$
is not suitable for enabling a shear-current dynamo
except perhaps for high values of $\Rm$.
On the other hand, as demonstrated in \Sec{LSfields}, large-scale magnetic
fields are being generated.
Explaining this in terms of the shear--current effect is very questionable.
Therefore we ask now whether an incoherent
alpha--shear dynamo (Vishniac \& Brandenburg 1997) might play a role.
Another explanation would be an incoherent shear-current dynamo
that we discuss below in \Sec{incohalpsh}.

The possibility that a random $\alpha$ with zero mean can produce
magnetic fields was first discussed by Kraichnan (1976) and Moffatt (1978).
In the presence of shear, strong large-scale fields can
be generated (Vishniac \& Brandenburg 1997, Sokolov 1997, Silant'ev 2000,
Fedotov et al.\ 2006, Proctor 2007).
Consider an incoherent alpha-shear dynamo with a scalar $\alpha$ fluctuating around zero.
In the limit $k \ll k_{\rm f}$ and if $|\alpha| k \ll |S|$
the condition for mean fields growing on the average exponentially reads
\EQ
D_{\alpha S}=
\alpha_{\rm rms}|S|/(\eta_{\rm T}^2k^3)
>D_{\alpha S}^{\rm crit},
\label{Dcrit}
\EN
where $D_{\alpha S}^{\rm crit}\approx2.3$ for a white noise $\alpha$ effect;
see \App{IncoherentAlpha}.

\begin{figure}[t!]
\centering\includegraphics[width=\columnwidth]{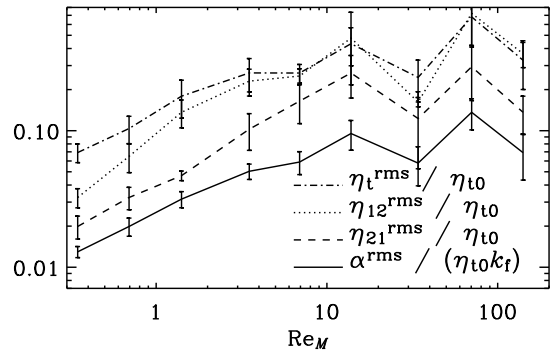}\caption{
Dependences of the rms values of the temporal fluctuations
$\alpha_{\rm rms}$ (normalized by $\eta_{\rm t0}k_{\rm f}$),
$\eta_{\rm t}^{\rm rms}$, $\eta_{21}^{\rm rms}$, and $\eta_{12}^{\rm rms}$
(normalized by $\eta_{\rm t0}$), on $\Rm$
for $\Rey=1.4$ and $\mbox{Sh}=-0.6$.
}\label{kinshear_Re14_alprms}\end{figure}

In a finite domain all mean--field coefficients show fluctuations,
and so $\alpha_{ij}$ must fluctuate about zero.
We may extend the test--field procedure for the determination of the
$\eta_{ij}$ such that it provides us the $\alpha_{ij}$, too.
When starting from (\ref{eq03}) instead of (\ref{eq10})
and using again the four test fields $\meanBB^{q {\rm c}}$ and $\meanBB^{q {\rm s}}$,  $q=1,2$,
we find
\begin{eqnarray}
\alpha_{i1} &=& B^{-1}\big( \meanemf^{1 {\rm c}}_i \cos kz + \meanemf^{1 {\rm s}}_i \sin kz \big)
\nonumber\\
\alpha_{i2} &=& B^{-1}\big( \meanemf^{2 {\rm c}}_i \cos kz + \meanemf^{2 {\rm s}}_i \sin kz \big) \, , \quad
    i = 1, 2\, ,
\label{eq19alp}
\end{eqnarray}
together with the relations (\ref{eq19}) for $\eta_{i1}$ and $\eta_{i2}$
(see also Brandenburg 2005b).

In contrast to the considerations in \Sec{Results}
we consider now the mean-field coefficients, as obtained from the test field calculations,
after averaging over $z$, but not over $t$.
Then the $\alpha_{ij}$ consist of fluctuations around a zero mean, that is, we have an incoherent
$\alpha$ effect.
(Without the averaging over $z$ the fluctuations would be even bigger.)
In the case of fluctuations of $\eta_{21}$ and $\eta_{12}$
we speak analogously about an incoherent shear--current effect.

We have calculated the rms values of the temporal fluctuations
of the $\alpha_{ij}$ and $\eta_{ij}$ which are denoted by
$\alpha_{ij}^{\rm rms}$ and $\eta_{ij}^{\rm rms}$, respectively.
We have also taken the averages over all four components of
$\alpha_{ij}^{\rm rms}$ and over the
two diagonal components of $\eta_{ij}^{\rm rms}$ and denoted them by $\alpha^{\rm rms}$ and $\eta_{\rm t}^{\rm rms}$,
respectively.
\FFig{kinshear_Re14_alprms} shows these quantities along with $\eta_{12}^{\rm rms}$
and $\eta_{21}^{\rm rms}$ for a small Reynolds number
and moderate shear as functions of $\Rm$. They are all mildly growing.

\begin{figure}[t!]
\centering\includegraphics[width=\columnwidth]{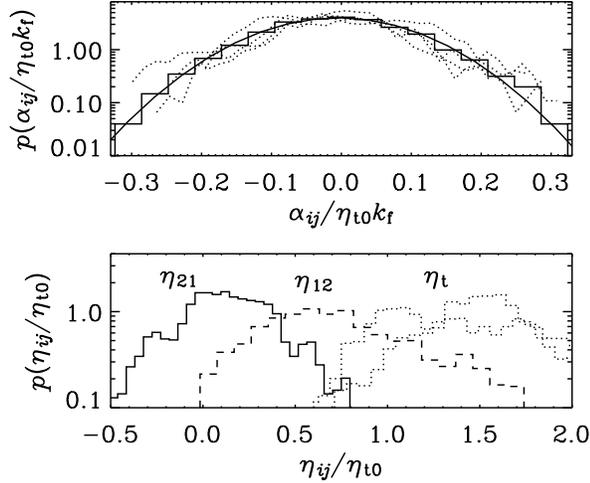}\caption{
Probability density functions (PDFs) of $\alpha_{ij}$ ({\it top}) and
$\eta_{ij}$ ({\it bottom}) for $\Rm=14$, $\Rey=1.4$, and $\mbox{Sh}=-0.6$.
The PDFs for the different components of $\alpha_{ij}$ ({\it dotted lines})
are close together; their average is given by the solid staircase line
and compared with a Gaussian fit.
The PDF of $\eta_{21}$ ({\it solid line}) is around zero while those of
$\eta_{12}$, $\eta_{11}$, and $\eta_{22}$ are not.
(The latter two are simply denoted by $\eta_{\rm t}$.)
}\label{pcomp_alphaeta_pdf}\end{figure}

In \Fig{pcomp_alphaeta_pdf} we show that the probability density functions
of $\alpha_{ij}$ and $\eta_{ij}$
for a run with $\Rm=14$, $\Rey=1.4$ and $\mbox{Sh}=-0.6$
are approximately Gaussian.
In order to improve the statistics we have, in addition, averaged
the results for all four components of $\alpha_{ij}$.
The result is similar to those for the individual components.
The diagonal components of $\eta_{ij}$ are distributed around finite
averages, and $\eta_{21}$ is distributed around a positive, but small value.

\subsection{An incoherent shear--current dynamo?}
\label{incohalpsh}

Yousef et al.\ (2007) have reported large-scale dynamo action
at low Reynolds numbers ($\Rey=\Rm=5$) for weak shear
[$\mbox{Sh}<(3\pi)^{-1}$] in tall boxes so that the smallest
wavenumber in the $z$ direction, $k_{1z}$,
can be up to 128 times smaller than those
in the other two directions.
They discuss in more detail the case where it is 16 times smaller,
i.e.\ $k_{1z}=k_1/16$.
Using small Reynolds numbers has the advantage that small-scale
dynamo action is then impossible.

\begin{figure}[t!]
\centering\includegraphics[width=\columnwidth]{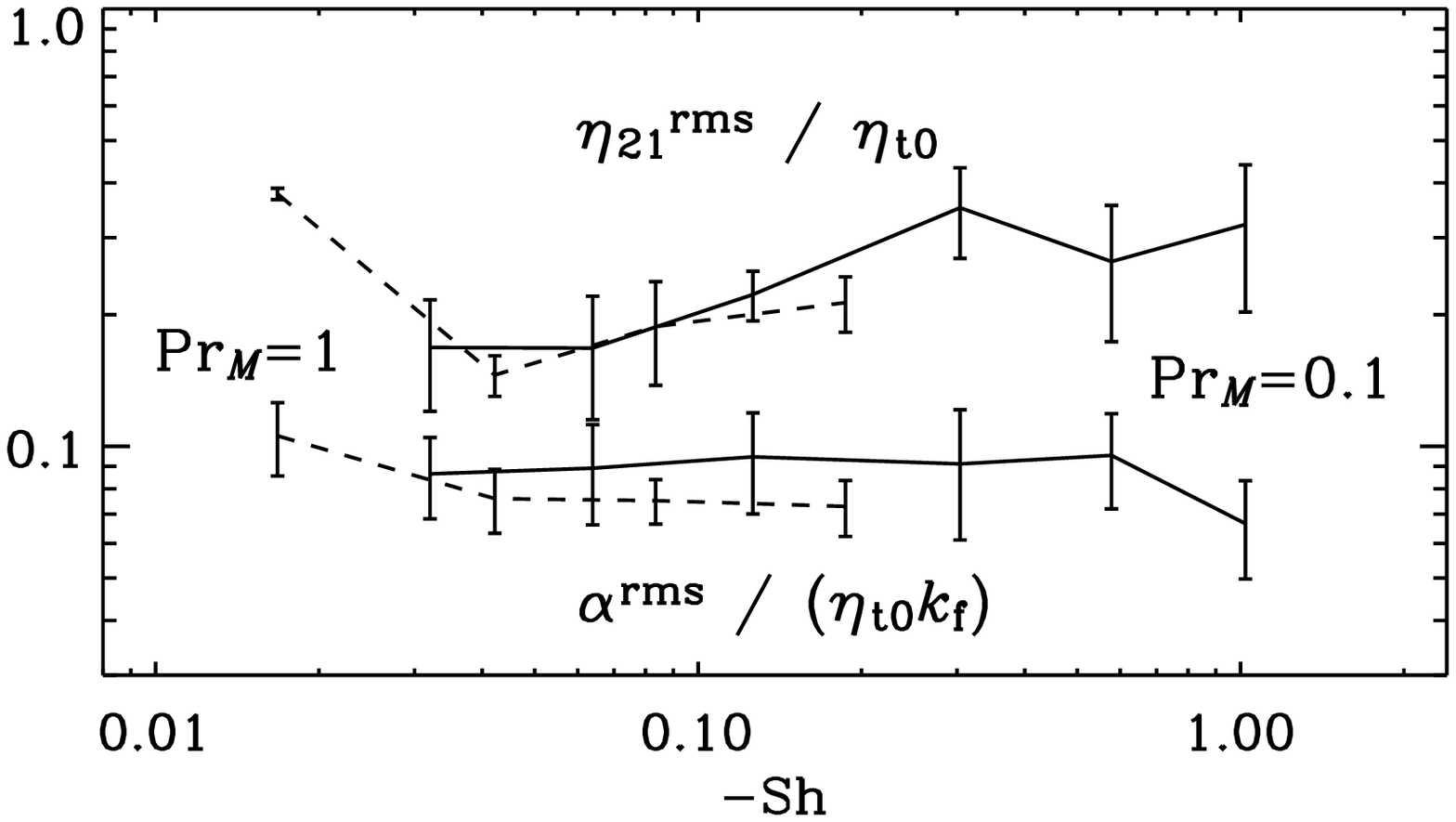}\caption{
Dependence of the rms values of the temporal fluctuations
$\alpha^{\rm rms}$ (normalized by $\eta_{\rm t0}k_{\rm f}$) and
$\eta_{21}^{\rm rms}$ (normalized by $\eta_{\rm t0}$), on $\mbox{Sh}$
for models with $\Pm=1$ and $\Rey=5$ ({\it dashed lines})
compared with the models shown in \Fig{kinshear_vs_shear} with
$\Pm=0.1$ and $\Rm=14$ ({\it solid lines}).
}\label{kinshear_vs_shear_alprms}\end{figure}

We have analyzed similar cases ($\Rey=5$ with $\Pm=1$ and $0.01<-\mbox{Sh}<0.3$)
using however cubic domains of size $L^3$, so $k_{1z}=k_1$,
and a forcing with $k_{\rm f}/k_1$ equal to 5 instead of 3.
It turns out that the value of the crucial coefficient
$\eta_{21}$ fluctuates around zero.
This is also plausible from \Fig{kinshear_vs_shear}
(even if it does not apply to $\Pm=1$).
We must therefore conclude that the (coherent) shear--current effect
cannot explain the generation of the mean magnetic field found
by Yousef et al.\ (2007).
With respect to the incoherent effects it can be seen from
\Fig{kinshear_vs_shear_alprms} that the values of
$\alpha^{\rm rms}$ and $\eta_{21}^{\rm rms}$ are more or less the same
for different $\Pm$, $\Rm$, and $\mbox{Sh}$.

Let us consider the instantaneous values of the growth rate
$\lambda$, as calculated from \Eq{lampm} with the fluctuating $\eta_{ij}$.
If $k=k_1$, we get always negative $\lambda$.
However, for the smaller value $k=k_1/16$,
appropriate for the model of Yousef et al.\ (2007),
it is possible to have large positive
$\lambda$ during extended periods of time.
Although $\meanBB$ can be amplified during those episodes it must decay during
episodes with the reversed sign of $\eta_{21}$, and it is not certain
from this competition whether a dynamo powered by the incoherent shear-current effect may result.

\begin{figure}[t!]
\centering\includegraphics[width=\columnwidth]{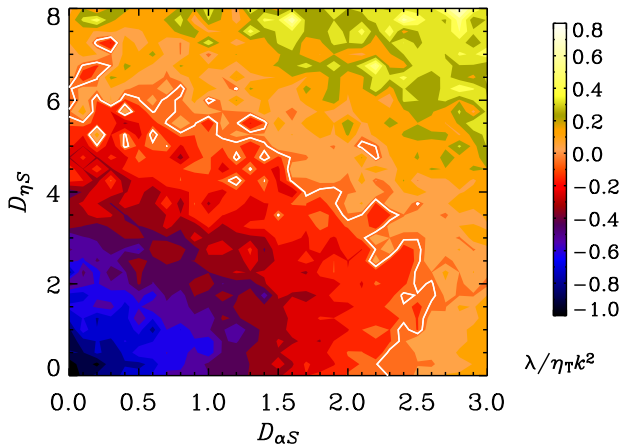}\caption{
Contour plot showing the normalized growth rate as a function of the
dynamo numbers for the incoherent alpha--shear and shear--current dynamos
from a numerical solution of the model described in Sect.~\ref{incohalpsh}.
The zero line is given in white.
}\label{psummary}\end{figure}

So far we ignored the possibility of an incoherent alpha--shear
dynamo that must work at the same time.
In order to assess the relative importance of the two incoherent effects
we have considered a simple model
with random $\alpha$ and $\eta$ tensors that are delta-correlated in time.
(Delta-correlated noise is the simplest model; a more realistic case
would be to assume colored noise with a finite correlation time.)
The model is explained in Appendix \ref{IncoherentAlpha}.
Under the assumptions $|\alpha_{11}|\ll|S|/k$
and $|\alpha_{12}|,|\alpha_{21}|\ll\eta_{\rm T}k$,
as well as $|\eta_{12}|\ll |S|/k^2$
and $|\epsilon| \ll |\eta_{\rm T}|$, its
governing parameters are the two dynamo numbers for the incoherent effects,
\EQ
D_{\eta S}=C_{\eta}C_S,\quad
D_{\alpha S}=C_{\alpha}C_S,
\EN
which can be expressed in terms of the three quantities
\EQ
C_S={|S|\over\eta_{\rm T}k^2},\quad
C_\alpha={\alpha_{22}^{\rm rms}\over\eta_{\rm T}k},\quad
C_\eta={\eta_{21}^{\rm rms}\over\eta_{\rm T}}\, . \label{Cs}
\EN
In \Fig{psummary} we give a contour plot of the normalized growth rate
as a function of the two dynamo numbers, $D_{\alpha S}$ and $D_{\eta S}$.
For small values of $D_{\eta S}$, the incoherent shear--current effect
has a slightly adverse effect on the dynamo, but for larger values it lowers
the critical value of $D_{\alpha S}$ significantly.
For $D_{\alpha S}=0$ even a purely incoherent shear--current dynamo is
possible if $D_{\eta S}\gtrsim 6.5$.

In interpreting simulations, we focus on domains whose smallest
finite wavenumber in the $z$ direction is $k_{1z}$.
In view of the already discussed dynamo found by Yousef et al.\ (2007)
we employ the data for $\alpha^{\rm rms}_{ij}$ and $\eta^{\rm rms}_{ij}$ originating from our
aforementioned calculations of similar cases to derive growth rates with the help of \Fig{psummary}.
With $k=k_{1z}= k_1/16$ we find first $C_S\approx40$, and from
\Figs{kinshear_Pm20}{kinshear_vs_shear_alprms}
we have $C_\eta\approx0.1$ and $C_\alpha\approx0.1$ (using $k_{\rm f}/k_1=3$),
so that $D_{\eta S}\approx4$ and $D_{\alpha S}\approx4$.
This suggests that the incoherent shear--current dynamo is
subcritical, while the incoherent alpha--shear dynamo is
supercritical.
Therefore an incoherent alpha--shear dynamo seems to be a plausible explanation.
This explanation is further supported by our finding that for constant
rms values the growth rate is in good approximation a linear function
of $S$ -- just as observed by Yousef and coworkers.

An explanation in terms of the incoherent $\alpha$ effect is also suited
for the nonhelical dynamo of Brandenburg (2005a),
where $C_S\approx25$, $C_\eta\approx0.15$ and $C_\alpha\approx0.2$
for the appropriate value of the magnetic Reynolds number, $\Rm=80$,
so that $D_{\eta S}\approx4$ and $D_{\alpha S}\approx5$.
For this model we used $k_{\rm f}/k_1=5$, although the shear profile
is here more complicated.

Finally, let us return to
the cases of dynamo action considered in \Figs{pmeanfield_w256a}{pmeanfield_w256b}.
If we assume that the negative values of $\eta_{21}$ seen in
\Fig{kinshear_Pm20} for $\Rm>100$ are real, we
have to ask for the relative importance of the regular (coherent)
shear--current effect and the two incoherent effects.
For both dynamo cases in \Figs{pmeanfield_w256a}{pmeanfield_w256b}
the values of $C_S$ are $\sim 30$ (but uncertain because the dynamo-generated
field quenches the value of $\eta_{\rm t}$), so the three dynamo numbers
would be $D_{\eta S}\approx1.5$ for the regular
shear--current effect, $D_{\eta S}=1.5\ldots3$ for the incoherent one and
$D_{\alpha S}=3\ldots6$, respectively, where we have used
$C_\eta=0.05\ldots0.1$ and $C_\alpha=0.1\ldots0.2$
(cf.\ \Figsss{kinshear_Pm20}{kinshear_Re14_alprms}{kinshear_vs_shear_alprms}).
The values of $D_{\alpha S}$ and $D_{\eta S}$ could be somewhat smaller
if one takes into account that the level of fluctuations
is smaller for $k_{\rm f}/k_{1z}=10$ instead of 5.
We recall that the
corresponding critical values are 1, 6.5 and 2.3, respectively.
Hence, with respect to the regular shear-current dynamo this case is
only slightly supercritical, but subcritical with respect to the
incoherent shear-current effect and supercritical with respect to the
incoherent alpha--shear dynamo.
By inspection of the values of the growth rate it is possible to infer safely that this situation is
dominated by the incoherent alpha effect. The incoherent shear-current effect has a weakly adverse influence whereas 
its regular counterpart clearly supports dynamo action.

\section{Conclusions}

The present work has demonstrated that the test field method
provides a robust means of determining all components
of the turbulent magnetic diffusivity tensor that are
relevant for mean fields depending only on $z$ and $t$.
Both rotating and  weakly shearing turbulence are studied.
In either case the
diagonal components of the turbulent diffusivity tensor
are about equal to each other.
Shear slightly enhances the turbulent magnetic diffusivity while
rotation quenches it.
In the presence of rotation, the $\OO\times\meanJJ$ effect
occurs, which is described by the off-diagonal components of the turbulent magnetic diffusivity tensor.
Shear leads to the shear--current effect, again described by off-diagonal components of this tensor.
In both cases the results are consistent with those found in the framework of the second-order correlation
approximation.

The possibility of the so-called shear-current dynamo has been scrutinized.
It depends crucially on the sign of the
component $\eta_{21}$ of the magnetic diffusivity tensor. It turns out that,
within the ranges of parameters considered,
its sign is in general not suited for driving a dynamo based on this effect,
with a possible exception at large magnetic Reynolds numbers.
In this way the analytic results found in the second--order correlation approximation
for incompressible fluids (R\"udiger \& Kitchatinov 2006; R\"adler \& Stepanov 2006)
are confirmed and generalized.

Direct numerical simulations are presented which exhibit growing mean magnetic fields in shear flow
turbulence. An interpretation as a (coherent) shear-current dynamo is hardly possible.
Instead, it is argued that it can be explained by an incoherent alpha--shear dynamo.
The incoherent shear--current effect has also been determined, but it is found to be
less important.

\acknowledgements
A great deal of the work presented here was done during stays of K.-H.\ R.\ and M.\ R.\ at NORDITA.
They are grateful for the hospitality.
A.\ B.\ thanks the Kavli Institute for Theoretical Physics for hospitality.
This research was supported in part by the National Science
Foundation under grant PHY05-51164.
P.\ J.\ K.\ acknowledges financial support from the Helsingin Sanomat foundation.
We thank the Centers for Scientific Computing in Denmark (DCSC), Finland (CSC),
and Sweden (PDC) for the allocation of computing resources.

\appendix
\section{Concerning $\alpha_{\lowercase{ij}}=0$}
\label{JustificationAlp0}

In the case of rotation the tensorial structure of $\alpha_{ij}$ must agree
with that of $\eta_{ij}$ given in (\ref{eq04}), that is,
\EQ
\alpha_{ij} = \alpha_0 \delta_{ij} + \alpha_1 \epsilon_{ijk} \hat{\Omega}_k
    + \alpha_2 \hat{\Omega}_i \hat{\Omega}_j \, .
\label{A01}
\EN
Since $\alpha_{ij}$ is a pseudo--tensor and $\OO$ an axial vector,
the coefficients $\alpha_0$, $\alpha_1$ and $\alpha_2$ must be pseudo--scalars.
Under our assumptions, however, no pseudo--scalars can be constructed.
So we have to conclude that $\alpha_{ij} = 0$.

In the case of shear we may argue analogously.
Referring to (\ref{eq06}) and the subsequent explanations we have then
\EQ
\alpha_{ij} = \alpha_0 \delta_{ij} + \alpha_1 g_i g_j + \alpha_2 h_i h_j
    + \alpha_3 g_i h_j + \alpha_4 g_j h_i
\label{A03}
\EN
with $\alpha_0$, $\alpha_1$ $\cdots$ $\alpha_4$ being pseudo--scalars.
Again, it is impossible to construct pseudo--scalars.
Thus, we have again $\alpha_{ij} = 0$.
Of course, the situation would be different
if the shear provided (large-scale) kinetic helicity,
as then the pseudo-scalar $\UU^S \cdot \curl \UU^S$ would be available.

\section{Comparison with results of the second--order correlation approximation}
\label{SOCA}

In a paper by R\"adler \& Stepanov (2006, referred to as RS06 in the following)
the mean electromotive force has been calculated in the second--order correlation approximation
for generally inhomogeneous turbulence in an incompressible rotating fluid
showing a position--dependent mean motion.
In this context the second--order correlation approximation was understood
as the neglect of higher--order terms
in the induction equation as well as in the momentum balance.
Both the Coriolis force and derivatives of the mean velocity were assumed to be small enough
so that the mean electromotive force is linear in the angular velocity $\OO$
and the gradient tensor of $\meanUU$.
Detailed results were obtained for a special correlation function of the background turbulence.

Let us apply the results to the situations considered in the present paper.
In the case of rotation without shear we obtain
\EQ
\frac{\delta}{\eta_{\rm t}} =
\frac{1}{4}\sqrt{\frac{\pi}{2}}\,
\Co \, \Rm \, (\lambda_{\rm c} k_{\rm f})^2
    \sqrt{q} \, \delta^0 (q, \Pm)
\label{B01}
\EN
with $\Co$, $\Rm$ and $\Pm$ as defined above, $q = \lambda_{\rm c}^2 / \eta \tau_{\rm c}$,
and $\lambda_{\rm c}$ and $\tau_{\rm c}$ being correlation length and time, respectively.
When introducing the Strouhal number $\mbox{St} = u_{\rm rms} k_{\rm f} \tau_{\rm c}$,
we have $q = \Rm / \mbox{St}$.
It seems plausible to assume that $\lambda_{\rm c} k_{\rm f} \approx 2 \pi$.
The function $\delta^0$ can be calculated according to
$\delta^0 = [\delta^{0 (\Omega)}(q, \Pm) + \kappa^{0 (\Omega)}(q, \Pm)] / 2 \beta^{0 (0)}(q)$
from the functions $\delta^{0 (\Omega)}$, $\kappa^{0 (\Omega)}$ and $\beta^{0 (0)}$
defined and plotted in RS06.
It turns out that $\delta^0$
is never negative and approaches unity if $\Pm = 1$ and $q \to 0$.
Of course, we have  $\delta / \eta_{\rm T} = (\delta / \eta_{\rm t}) (\eta_{\rm t} / \eta_{\rm T})$.
The factor $\eta_{\rm t} / \eta_{\rm T}$ depends on $\Rm$, $\lambda_{\rm c} k_{\rm f}$
and $\beta^{0 (0)}(q)$.
It satisfies $0 \leq \eta_{\rm t} / \eta_{\rm T} < 1$ and approaches unity as $\Rm \to \infty$.

Clearly \eq{B01} and the results reported in \Sec{EffectRot} agree in the sign of $\delta$.
Although these results do not really confirm the linearity of $\delta$ in $\Co$,
which is suggested by \eq{B01}, they are not in conflict with that, see \Fig{kinrot_vs_Co}.
A further comparison of results is difficult because of, e.g., the
not exactly known value of $\lambda_{\rm c}$
and the errors of the data presented above.

Proceeding to the case of shear without rotation we note first
that, due to the aforementioned assumption on the linearity in the mean--velocity gradient,
that is in $S$, both $\kappa_{11}$ and $\kappa_{22}$ are equal to zero.
Furthermore, we have
\EQ
\frac{\eta_{12}}{\eta_{\rm t}} = - \frac{3}{5} \mbox{Sh} \, \Rm (\lambda_{\rm c} k_{\rm f})^2
    \eta_{12}^0 (q, \Pm) \, , \quad
    \frac{\eta_{21}}{\eta_{\rm t}} = - \frac{3}{5} \mbox{Sh} \, \Rm (\lambda_{\rm c} k_{\rm f})^2
    \eta_{21}^0 (q, \Pm) \, .
\label{B05}
\EN
Here $\eta^0_{12} = \half [\eta^{0 (D)} (q, \Pm) + \eta^{0 (W)} (q, \Pm)]$
and $\eta^0_{21} = \half [\eta^{0 (D)} (q, \Pm) - \eta^{0 (W)} (q, \Pm)]$,
where $\eta^{0 (D)} = [13 \kappa^{0 (D)}(q, \Pm) - 7 \beta^{0 (D)} (q, \Pm)]/ 6 \beta^{0 (0)}(q)$
and $\eta^{0 (W)} = [5 \delta^{0 (W)}(q, \Pm) + \kappa^{0 (D)} (q, \Pm)]/ 6 \beta^{0 (0)}(q)$,
with the functions $\kappa^{0 (D)}$, $\beta^{0 (D)}$, $\delta^{0 (W)}$, $\kappa^{0 (W)}$
and $\beta^{0 (0)}$ of RS06.
The quantities $\eta^0_{12}$ and $\eta^0_{21}$ approach unity and zero, respectively,
if $\Pm = 1$ and $q \to 0$.
We note that $- \eta_{12} / S$ and $\eta_{21} / S$ coincide with the quantities $\delta'$ and $\delta$
introduced in Appendix D of RS06, respectively.
It has been shown there that this $\delta$ (different from that considered above)
cannot take negative values. This applies then to $\eta_{21}$, too.

Being aware that the second--order approximation applies only for $\Rm$ that is not too large, we may state
that \eq{B05} and the numerical results reported in \Sec{EffectShear} agree in the sign of $\eta_{21}$.
The possible deviation in \Fig{kinshear_Pm20} is outside the validity range of this approximation.
The linearity of $\eta_{12}$ and $\eta_{21}$ in $\mbox{Sh}$ indicated in \eq{B05} is well confirmed
by the numerical results; see \Fig{kinshear_vs_shear}.
Again, further comparison of the results is, for the reasons mentioned above, rather difficult
but no striking disagreement has been found.

\begin{figure}[t!]
\centering\includegraphics[width=\columnwidth]{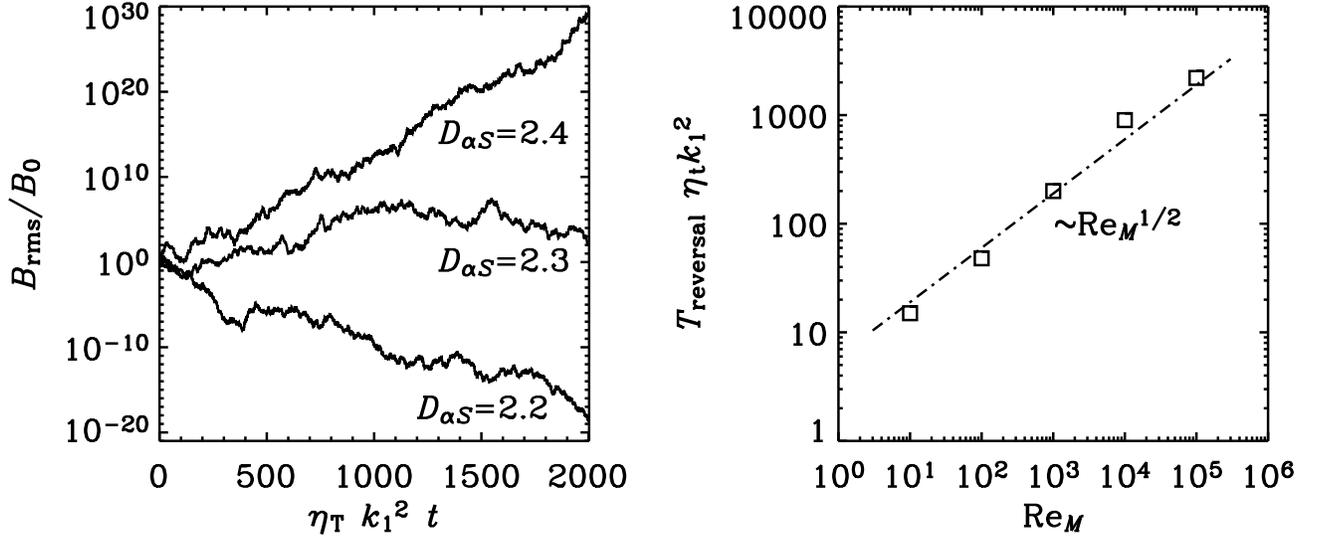}\caption{
Results for an incoherent alpha-shear dynamo ($D_{\eta S}=0$).
{\it Left}: Temporal evolution curves of $B_{\rm rms}$
showing that the critical value of $D_{\alpha S}$ is around 2.3.
Nonlinearity is here ignored and $B_{\rm rms}$ is scaled by its
initial value, $B_0$.
{\it Right}: With additional dynamical quenching, $D_{\alpha S}=50$
and $k_{\rm f}/k_1=5$.
Note that the typical time between reversals increases with $\Rm$
approximately like $\Rm^{1/2}$.
}\label{incoh_alpha}\end{figure}

\section{Incoherent alpha--shear and shear--current dynamos}
\label{IncoherentAlpha}

We calculate numerically solutions of the dynamo equation
with incoherent alpha and shear--current effects in unbounded space.
It reads
\EQ
{\DDD\meanAA\over\DDD t}=-S\meanA_y\xxx+\meanEMF-\eta\meanJJ,
\label{dAtildedt}
\EN
where
\EQ
\meanemf_i=\alpha_{ij}(t)\meanB_j-\eta_{ij}(t)\meanJ_j \,.
\EN
The $\alpha$ and $\eta$ tensors are delta-correlated in time with
\EQ
\bra{\alpha_{ij}(t)\,\alpha_{ij}(t')}=
\left(\alpha_{ij}^{\rm rms}\right)^2\delta(t-t'),
\quad\bra{\alpha_{ij}}=0,
\label{alptt1}
\EN
\EQ
\bra{\eta_{ij}(t)\,\eta_{ij}(t')}-\bra{\eta_{ij}}^2=
\left(\eta_{ij}^{\rm rms}\right)^2\delta(t-t'),
\label{etatt1}
\EN
where no summation over double-indices is assumed and $\bra{\ldots}$
means here a temporal or ensemble average.
For solving \eq{dAtildedt} we use the ansatz
\EQ
\meanAA(z,t)=\tilde{\AAA}(t)\exp(\ii kz)
\label{one-mode}
\EN
with an arbitrary, but fixed wavenumber $k$ and employ
a third-order Runge-Kutta time stepping scheme.
At each time step of length $\delta t$, the fluctuations of $\alpha_{ij}$
and $\eta_{ij}$ are taken as random numbers from a Gaussian distribution
and scaled by $1/\!\sqrt{\delta t}$ so that \Eqs{alptt1}{etatt1} hold.

We recall that, if $\alpha_{ij}= \alpha \delta_{ij}$ with
$\alpha=\const\ll S/k$, the critical value of the dynamo
number $D_{\alpha S}$ as defined in \Eq{Dcrit} but
with $\alpha^{\rm rms}$ replaced by $\alpha$,
is $D_{\alpha S}^{\rm crit}=2$
(e.g.\ Brandenburg \& Subramanian 2005a).
In the case of a pure incoherent alpha--shear dynamo,
i.e.\ $D_{\eta S}=0$, it is found that $D_{\alpha S}^{\rm crit}\approx2.3$ (see
the left panel of \Fig{incoh_alpha}).
On the other hand, for $D_{\alpha S}=0$ we have
$D_{\eta S}^{\rm crit}\approx6.5$.

There are reversals on a typical timescale of about one diffusion time.
However, this time can increase significantly if magnetic helicity
conservation (appropriate for a closed domain) is taken into account
(Field \& Blackman 2002, Blackman \& Brandenburg 2002, Subramanian 2002).
This means that the $\alpha$ effect has to be amended by an additional
term that results from the current helicity produced by the
dynamo.
We assume again $\alpha_{ij}= \alpha \delta_{ij}$ 
and a non-fluctuating $\eta_{ij}= \eta_{\rm t} \delta_{ij}$, further
\EQ
\alpha(t)=\alpha_{\rm K}(t)+\alpha_{\rm M}(t),
\EN
where $\alpha_{\rm K}(t)$ is stochastic, just like $\alpha_{ij}(t)$
in \Eq{alptt1}, and $\alpha_{\rm M}(t)$ obeys the differential equation
\EQ
{\dd\alpha_{\rm M}\over\dd t}=-2\eta_{\rm t}k_{\rm f}^2
\left({|\tilde{\emf}^*\cdot\tilde{\BB}|\over B_{\rm eq}^2}
+{\alpha_{\rm M}\over\Rm}\right)
\label{dalpdt}
\EN
with $\tilde{\BB}$ and $\tilde{\emf}$ defined analogously to $\tilde{\AAA}$ in \eq{one-mode}.
The dynamo number $D_{\alpha S}$ is now defined with respect to
$\alpha_{\rm K}^{\rm rms}$.
\EEq{dalpdt} is solved simultaneously with \Eq{dAtildedt} using
the aforementioned time stepping scheme.
As here a nonlinearity is introduced, the ansatz (\ref{one-mode}) has now to be understood
as a one-mode truncation.
The model calculations show that the timescale for reversals
increases proportional to $\Rm^{1/2}$ (see the right panel of
\Fig{incoh_alpha}).

\end{document}